\RequirePackage{fix-cm}
\documentclass{svjour3}                     
\usepackage{amssymb}
\usepackage{algorithm}
\usepackage{latexsym}
\usepackage{mathtools}
\smartqed  
\usepackage{booktabs}
\usepackage{enumerate}
\usepackage{graphicx}
\usepackage{tikz}
\usepackage{lmodern}
\usepackage{microtype}
\usepackage{subfigure}
\DeclarePairedDelimiter{\norm}{\lVert}{\rVert}

\begin{document}

\title{Newton-Type Optimal Thresholding Algorithms for Sparse Optimization Problems
\thanks{The work was founded by the Natural Science Foundation of China (NSFC) under the grant 12071307. }
}

\titlerunning{Newton-Type Optimal $k$-Thresholding}        

\author{    Nan Meng   \and Yun-Bin Zhao   }


\institute{Nan Meng\at
              School of Mathematics, University of Birmingham, Edgbaston, Birmingham, B15 2TT, United Kingdom \\
              \email{nxm563@bham.ac.uk}           
           \and
           Yun-Bin Zhao \at
              Shenzhen Research Institute of Big Data, Chinese University of Hong Kong, Shenzhen, China \\
              \email{yunbinzhao@cuhk.edu.cn}
}

\date{Received: date / Accepted: date}

\maketitle

\begin{abstract}
    Sparse signals can be possibly reconstructed by an algorithm which merges a traditional nonlinear optimization method and a certain thresholding technique.
    Different from existing thresholding methods, a novel thresholding technique referred to as  the optimal $k$-thresholding was recently proposed by  Zhao [SIAM J Optim, 30(1), pp. 31-55, 2020].
    This technique simultaneously performs the minimization of an error metric for the problem and thresholding of the iterates generated by the classic gradient method.
    In this paper, we propose the so-called Newton-type optimal $k$-thresholding (NTOT) algorithm which is motivated by the appreciable performance of both Newton-type methods and the optimal $k$-thresholding technique for signal recovery.
    The guaranteed performance (including convergence) of the proposed algorithms are shown in terms of suitable choices of the algorithmic parameters and the restricted isometry property (RIP) of the sensing matrix which has been widely used in the analysis of compressive sensing algorithms.
    The simulation results based on synthetic signals indicate that the proposed algorithms are stable and efficient for signal recovery.
\end{abstract}

\section{Introduction}
The sparse optimization problem arises naturally from  a wide range of practical scienarios such as compressed sensing \cite{candes2006robust,eldar2012compressed,foucart2013mathematical,zhao2018sparse}, signal and image processing \cite{Boche19,Maio19,elad2010sparse}, pattern recognition \cite{patel2011sparse} and wireless communications \cite{Choi17}, to name a few.
The typical problem of signal recovery via compressed sensing can be formulated as the following sparse optimization problem:
\begin{equation} \label{prob}
	\min _{x}\left\{\|y-A x\|_{2}^{2}: ~ \|x\|_{0} \leq k\right\} ,
\end{equation}
where $ k$ is a given integer number reflecting the sparsity level of the target signal $x^*,$  $A \in \mathbb{R}^{m\times n}$ is a measurement matrix with $m \ll n,$ $\|x\|_0$ is the so-called  $\ell_0$-norm counting the nonzeros of the vector $x,$ and $y$ are the acquired  measurements of the signal $x^*$ to recover. The  vector $y$ is ususally represented as $ y= Ax^*+\eta, $ where $\eta$ denotes a noise vector.

Developing effective algorithms for the model (\ref{prob}) is fundamentally important in signal recovery.
At the current stage of development, the main algorithms for solving sparse optimization problems can be categorized  into several classes: convex optimization, heuristic algorithms, thresholding algorithms, and Bayes methods. The typical convex optimization methods include $\ell_1$-minimization \cite{chen2001atomic,candes2005decoding}, reweighted $\ell_1$-minimization \cite{candes2008enhancing,zhao2012reweighted}, and dual-density-based reweighted $\ell_1$-minimization \cite{zhao2015Michal,zhao2017constructing,zhao2018sparse}.
The widely used heuristic algorithms include orthogonal matching pursuit (OMP) \cite{tropp2007signal,needell2010signal}, subspace pursuit (SP) \cite{dai2009subspace}, and compressive sampling matching pursuit (CoSaMP) \cite{needell2009cosamp,satpathi2017on}.
Depending on thresholding strategies, the thresholding methods can be roughly classified as soft-thresholding \cite{donoho1995denoising,fornasier2008iterative}, hard-thresholding   (e.g., \cite{blumensath2009iterative,blumensath2010normalized,blumensath2012accelerated,bouchot2016hard,nan2020newton}), and the so-called optimal thresholding methods \cite{zhao2020optimal,zhao2020analysis}.

The hard thresholding is the simplest thresholding approach used to generate iterates satisfying the constraint of the problem (\ref{prob}).
Throughout the paper, we use ${\cal H}_k (\cdot)$ to denote the hard thresholding operator which retains the largest $k$ magnitudes of a vector and zeroes out the others.  The following iterative hard thresholding (IHT) scheme
\begin{equation*} \label{eq1-3}
	x^{\text{p+1}}={\cal H}_k \left( x^p +\lambda A^{T}\left(y-A x^p \right) \right),
\end{equation*}
where $\lambda > 0$ is a stepsize, was first studied in \cite{blumensath2008iterative,blumensath2009iterative}. Incorporating a pursuit step (least-squares step) into IHT yields the hard thresholding pursuit (HTP) \cite{foucart2011hard,bouchot2016hard}, and when $ \lambda $ is replaced by an adaptive stepsize similar to the one used in traditional conjugate methods, it leads to the so-called  normalized iterative hard thresholding (NIHT) algorithms in \cite{blumensath2010normalized,tanner2013normalized}.
The theoretical performance of these algorithms can be analyzed in terms of the restricted isometry property (RIP)(see, e.g., \cite{blumensath2008iterative,blumensath2009iterative,foucart2013mathematical}).

On the other hand, the search direction  $ A^T (y-Ax^p) $ of the above-mentioned algorithm is the negative gradient of the objective function of the problem (\ref{prob}). Such a search direction can be replaced by another direction provided that it is a descent direction of the objective function.
Thus an Newton-type direction was studied in \cite{zhou2019global,nan2020newton,zhou2020subspace}. The following iterative method is proposed and referred to as NSIHT in \cite{nan2020newton}:
\begin{equation} \label{eq1-4}
	x^{p+1}={\cal H}_k \left( x^p +\lambda\left(A^{T} A+\epsilon I\right)^{-1} A^{T}\left(y-A x^p \right) \right),
\end{equation}
where $\epsilon > 0$ is a parameter and $\lambda > 0$ is the stepsize.

However, as pointed out in \cite{zhao2020optimal,zhao2020analysis}, the weakness of the hard thresholding operator ${\cal H}_k (\cdot)$ is that when applied to a non-sparse iterate generated by the classic gradient method, it may cause a ascending value of the objective  of (\ref{prob}) at the thresholded vector, compared to the objective value at its unthresholded counterpart.
As a result, direct use of the hard thresholding operator to a non-sparse or non-compressible vector in the course of an algorithm may lead to significant numerical oscillation and divergence of the algorithm.
 To overcome such a drawback of hard thresholding operator, Zhao \cite{zhao2020optimal} proposed an optimal $k$-thresholding technique which performs thresholding and objective-value reduction simultaneously. The optimal $k$-threholding iterative scheme  in \cite{zhao2020optimal} can be simply stated as
 \begin{equation*}
	x^{p+1}={\cal Z}^{\#}_k \left( x^p +\lambda A^{T}\left(y-A x^p \right) \right),
\end{equation*}
where $\lambda$ remains a stepsize, and ${\cal Z}^{\#}_k (\cdot)$ is the so-called optimal $k$-thresholding operator.
Given a vector $u$, the thresholded vector ${\cal Z}^{\#}_k  (u) = u\otimes w^*$ (the Hadamard product of two vectors) where the vector $ w^*$ is the optimal solution to the following quadratic $0$-$1$ optimization problem:
\begin{equation*}
	w^* :=\textrm{arg}\min _{w}\left\{\|y-A(u \otimes w)\|_{2}^{2}:  ~ \mathbf{e}^{T} w=k, ~ w \in\{0,1\}^{n}\right\},
\end{equation*}
where $\mathbf{e} = (1, \dots, 1)^T \in \mathbb{R}^n  $ is the vector of ones, and $\{0,1\}^{n} $ denotes the set of $n$-dimensional $0$-$1$ vectors.
To avoid solving such a binary optimization problem, an alternative approach is to solve its convex relaxation which, as pointed out in \cite{zhao2020optimal,zhao2020analysis}, is the tightest convex relaxation of the above problem:
\begin{equation} \label{eq1-6}
	\widehat{w} :=\textrm{arg}\min _{w}\left\{\|y-A(u \otimes w)\|_{2}^{2}: ~ \mathbf{e}^{T} w=k, ~ 0 \le w \le \mathbf{e} \right\}.
\end{equation} Based on the convex relaxation of the operator ${\cal Z}^{\#}_k (\cdot) ,$  efficient algorithms called relaxed optimal $k$-thresholding algorithms (ROT) and its variants have been proposed and
investigated in \cite{zhao2020optimal,zhao2020analysis}.
Simulations demonstrate that this new framework of thresholding methods works efficiently, and it overcomes the drawback of the traditional hard thresholding operator.

Due to the aforementioned weakness of ${\cal H}_k$ which appears in the Newton-type iterative method (\ref{eq1-4}), it makes sense to consider a further improvement of the performance of such a method.
The purpose of this paper is to combine the optimal $k$-thresholding and Newton-type search direction in order to develop an algorithm that may alleviate or eliminate the drawback of hard thresholding operator and hence enhance the numerical performance of the Newton-type method \eqref{eq1-4}.
The proposed algorithms are called the Newton-type optimal $k$-thresholding (NTOT).
The convex relaxation versions of this algorithm are also studied in this paper, which are referred to as Newton-type relaxed optimal thresholding (NTROT) algorithms, and its enhanced version with a pursuit step (NTROTP for short).
The guaranteed performance and convergence of these algorithms are shown under the RIP assumption as well as suitable conditions imposed on the algorithmic parameters.
	
The paper is organized as follows.
The algorithms are described in Section 2.
The theoretical performances of the proposed algorithms in noisy settings are shown in Section 3.
The empirical results are  demonstrated in Section 4, which indicate that under appropriate choices of the parameter and stepsize the proposed algorithms  are efficient for signal reconstruction and their performances are comparable to a few existing methods.

\section{Algorithms}

Some notations will be used throughout the paper. Let $\mathbb{R}^{n}$ denote the $n$-dimensional Euclidean space, and $\mathbb{R}^{m \times n}$ denotes the set of $m \times n$ matrices.
For a vector $x \in \mathbb{R}^{n}$, the $\ell_2$-norm is defined as $\|x\|_{2}:=\sqrt{\sum_{i}^{n} x_{i}^{2}}.$
We use  $[N]$ to denote the set $\{1, \ldots, n\}$.
Given a set $\Omega \subseteq[N]$, $\overline{\Omega}:=[N] \backslash \Omega$ denotes the complement set of $\Omega.$
$x_{\Omega}$ denotes the vector obtained from $x$ by retaining the entries of $x$ indexed by $\Omega$ and zeroing out the ones indexed by $\overline{\Omega}.$
$A^{T}$ denotes the transpose of the matrix $A.$
Give a vector $u$, ${\cal L}_k(u)$ denotes the index set of the largest $k$ magnitudes of $u$.
Throughout the paper, a vector $x$ is said to be $k$-sparse if $\|x\|_0 \le k$.

Note that the gradient and Hessian of the function $f(x) = \frac{1}{2} \|y-A x\|_{2}^{2}$
are given as
\[ \nabla f(x)=-A^{T}(y-A x), \quad \nabla^{2} f(x)=A^{T} A . \]
For the problem (\ref{prob}), the Hessian  $A^{T} A $  is singular, and thus the classic Newton's method cannot be applied  to the function $f(x)$  directly.
Modifying the matrix by adding $\epsilon I$ leads to the nonsingular matrix $A^TA+\epsilon I ,$ where $\epsilon $ is a positive parameter and $I \in \mathbb{R}^{n \times n}$ is the identity matrix.
Then we immediately obtain the following Newton-type iterative method for the minimization of  $f(x):  $
\[ x^{p+1}=x^p+\lambda\left(A^{T} A+\epsilon I\right)^{-1} A^{T}(y-A x^p), \]
where $ \lambda $ is a stepsize.
Different from the approach \eqref{eq1-4}, we utilize the optimal $k$-thresholding operator instead of the hard thresholding operator to develop a Newton-type iterative algorithm, which is described as  Algorithm 1.

\begin{algorithm}
\caption{Newton-Type Optimal $k$-Thresholding (NTOT)}
\begin{itemize}
    \item Input: measurement matrix $A$, measurement vector $y$, sparsity level $k$, parameter $\epsilon>0$, and stepsize $\lambda$.
    Give an initial point $x^0 \in \mathbb{R}^n. $
    \item Iteration:
          \begin{align}
         u^p & = x^p + \lambda (A^TA + \epsilon I)^{-1} A^T(y-Ax^p) ,  \nonumber \\
         \overline{w}^p  & = \arg \min_w \{ \norm*{y-A(u^p \otimes w)}_2^2 : ~ \mathbf{e}^T w=k, ~ w \in \{ 0,1 \}^n \} ,   \tag{P$_1$} \\
          x^{p+1} & = u^p \otimes \overline{w}^p .  \nonumber
        \end{align}
    \item Output: $k$-sparse vector $\hat{x}$.
\end{itemize}
\end{algorithm}

Solving the $0$-$1$ problem (P$_1$) is generally expensive,  Zhao \cite{zhao2020optimal} suggests solving its tightest convex relaxation, i.e., the problem \eqref{eq1-6}. This results in the Newton-type relaxed optimal $k$-thresholding algorithm, which is described as Algorithm 2.

\begin{algorithm}
\caption{Newton-Type Relaxed Optimal $k$-Thresholding (NTROT)}
\begin{itemize}
    \item Input: measurement matrix $A$, measurement vector $y$, sparsity level $k$, parameter $\epsilon>0$, and stepsize $\lambda$.
    Give an initial point $x^0 \in \mathbb{R}^n. $
    \item Iteration:
         \begin{align}
          u^p &  = x^p + \lambda (A^TA + \epsilon I)^{-1} A^T(y-Ax^p), \nonumber \\
         w^p & = \arg \min_w \{ \norm*{y-A(u^p \otimes w)}_2^2 : ~ \mathbf{e}^T w=k, ~ 0 \le w \le \mathbf{e} \}, \tag{P$_2$} \\
         x^{p+1} & = {\cal H}_k (u^p \otimes w^p) .  \tag{P$_3$}
        \end{align}
    \item Output: $k$-sparse vector $\hat{x}$.
\end{itemize}
\end{algorithm}

If the step (P$_2$) generates a $0$-$1$ solution $w^p$, i.e., $ w^p$ is exactly a $ k$-sparse vector,  then $ u^p \otimes w^p$ is exactly $ k$-sparse, in which case the operator ${\cal H}_k$ in (P$_3$) is superfluous.
However, as the vector $ w^p$ may not necessarily be $k$-sparse, so $ {\cal H} _k$ is used in (P$_3$) to truncate the iterate so that it satisfies the constraint of the problem (\ref{prob}).
This is quite different from $ {\cal H}_k(u^p)$ that directly performs hard thresholding on $u^p$ which may not be sparse at all.
The vector $w^p$ are either $ k$-sparse or admits a compressible feature in which case performing hard thresholding on the resulting vector $ u^p\otimes w^p$ can avoid significant oscillation of the objective value of (\ref{prob}).

To further stabilize the NTROT, a pursuit step can be performed after solving the optimization problem (P$_2$).
This leads to the algorithm called NTROTP, which is the main algorithm concerned in this paper.

\begin{algorithm}
\caption{Newton-Type Relaxed Optimal $k$-Thresholding Pursuit (NTROTP)}
\begin{itemize}
    \item Input: measurement matrix $A$, measurement vector $y$, sparsity level $k$, parameter $\epsilon>0$, and stepsize $\lambda$.
    Give an initial point $x^0 \in \mathbb{R}^n. $
    \item Iteration:
         \begin{align}
         u^p &= x^p + \lambda (A^TA + \epsilon I)^{-1} A^T(y-Ax^p), \nonumber \\
          w^p &= \arg \min_w \{ \norm*{y-A(u^p \otimes w)}_2^2 : ~ \mathbf{e}^T w=k, ~ 0 \le w \le \mathbf{e} \}, \tag{P$_4$} \\
          S^{p+1} &= {\cal L}_k (u^p \otimes w^p),   \nonumber \\
          x^{p+1} &= {\arg\min}_{z} \{ \norm{y-Az}_2 : \operatorname{supp}(z) \subseteq S^{p+1} \} .  \tag{P$_5$}
        \end{align}
    \item Output: $k$-sparse vector $\hat{x} $.
\end{itemize}
\end{algorithm}
The step (P$_5$) is a  pursuit step at which a least-squares problem is solved on the support of the largest $k$ magnitudes of the vector $ u^p \otimes w^p. $
In the next section, we establish sufficient conditions for the guaranteed performance and convergence of the algorithms NTOT, NTROT and NTROTP.

\section{Theoretical Analysis}
Before going ahead, let us first recall the definition of restricted isometry constant (RIC).

\begin{definition}\cite{candes2005decoding}
The $q$-th order restricted isometry constant (RIC) $\delta_q$ of a matrix $A \in \mathbb{R}^{m \times n}$ is the smallest number $\delta_q \geq 0 $ such that
\[ \left(1-\delta_{q}\right)\|x\|_{2}^{2} \leq\|A x\|_{2}^{2} \leq\left(1+\delta_{q}\right)\|x\|_{2}^{2} \]
for all $q$-sparse vectors $x$, where $q$ is an integer number.
\end{definition}
If $ \delta_q<1$, we say that the  matrix $A$ satisfies the $q$-th order restricted isometry property (RIP).
It is well known that the random matrices including Bernoulli, Gaussian and more general subgaussian matrices may satisfy the  RIP of a certain order with an overwhelming probability \cite{candes2005decoding,candes2006robust,foucart2013mathematical}.

\subsection{Analysis of NTOT in noisy scenarios}

The following two lemmas are very helpful to show the main result in this section.
The first one was taken from \cite{nan2020newton} and the second one can be found in \cite{zhao2020optimal}.

\begin{lemma} \emph{\cite{nan2020newton}} \label{lem1}
Let $ A\in \mathbb{R}^{m\times n}$ with $ m\ll n$ be a measurement matrix.
Given a vector $u \in \mathbb{R}^{n}$ and an index set $\Omega \subset{[N]}$, if $(\epsilon, \lambda)$ is chosen such that $\epsilon > \sigma_1^2$ and $\lambda \le \epsilon + \sigma_m^2$, where $\sigma_1,\sigma_m$ are the largest and smallest singular values of the matrix $A,$ respectively, then one has
\[ \begin{aligned}
    & \left\|\left[ (I - \lambda (A^TA+\epsilon I)^{-1} A^TA ) u \right]_\Omega \right\|_2 \le
     ( \delta_t + \sigma_1^2-\frac{\lambda\sigma_1^2}{\epsilon+\sigma_1^2} ) \|u\|_2
\end{aligned} \]
provided that $|\Omega \cup \operatorname{supp}(u)| \le t$, where $ t$ is a certain integer number.
\end{lemma}

\begin{lemma} \emph{\cite{zhao2020optimal}} \label{lem2}
Let $y=A \hat{x} + \eta$ be the measurements of the $k$-sparse vector $\hat{x} \in \mathbb{R}^n$, and let $u \in \mathbb{R}^n$ be an arbitrary vector.
Let ${\cal Z}_k^{\#}(u) $ be the optimal $k$-thresholding vector of $u$.
Then for any $k$-sparse binary vector $\hat{w} \in \{0,1\}^n$ satisfying $\operatorname{supp}(\hat{x}) \subseteq \operatorname{supp}(\hat{w})$, one has
\begin{equation} \label{lem2-1}
	\norm*{{\cal Z}_k^{\#}(u) - \hat{x}}_2 \le \sqrt{\frac{1+\delta_k}{1-\delta_{2k}}} \norm*{(\hat{x} - u) \otimes \hat{w}}_2 + \frac{2}{\sqrt{1-\delta_{2k}}} \| \eta \|_2.
\end{equation}
\end{lemma}
The bound \eqref{lem2-1} follows directly from the proof of Theorem 4.3 in \cite{zhao2020optimal}.
In fact, the inequality \eqref{lem2-1} is obtained by combining the inequality (4.5) and the first inequality of (4.7) in \cite{zhao2020optimal}.

We now state and show the sufficient condition for the guaranteed performance of NTOT in noisy settings.

\begin{theorem}
    Let $y=Ax+\eta$ be the measurements of the signal $x \in \mathbb{R}^n$ with measurement error $ \eta.$
    Let $S={\cal L}_k(x)$ and $\sigma_1$ and $\sigma_m$ be, respectively, the largest and smallest singular values of the matrix $A \in \mathbb{R}^{m \times n}$.
    Suppose that the restricted isometry constant of $A$ satisfies
    $$ \delta_{2k} < 0.5349,  $$
and that $\epsilon$ is a given parameter satisfying
    \begin{equation} \label{LLL}
    	\epsilon > \max \left\{ \sigma_1^2, \left( \frac{\sigma_1^2 - \sigma_m^2}{\sqrt{\frac{1-\delta_{2k}}{1+\delta_k}} - \delta_{2k}} - 1 \right) \sigma_1^2 \right\} .
    \end{equation}
    If the stepsize $\lambda$ in NTOT is chosen such that
       \begin{equation} \label{LLLL}
       		\epsilon + \sigma_1^2 + \left( \delta_{2k} - \sqrt{\frac{1-\delta_{2k}}{1+\delta_k}} \right) \frac{\epsilon + \sigma_1^2}{\sigma_1^2} ~ < ~ \lambda ~ \le ~ \epsilon + \sigma_m^2,
        \end{equation}
    then the sequence $\{x^p\} $ generated by the NTOT satisfies that
    \begin{equation} \label{thm1result}
    	\left\|x^{p+1}-x_S\right\|_{2} \leq \rho\left\|x^{p}-x_{S}\right\|_{2}+\tau \left\|Ax_{\overline{S}} + \eta \right\|_{2},
    \end{equation}
    where
    \[ \rho = \sqrt{\frac{1+\delta_{k}}{1-\delta_{2 k}}}\left(\delta_{2 k}+\sigma_{1}^{2}-\frac{\lambda \sigma_{1}^{2}}{\epsilon+\sigma_{1}^{2}}\right) \]
    and
    \[ \tau = \frac{1}{\sqrt{1-\delta_{2 k}}} \left( \frac{\lambda \sigma_{1}^{2} \sqrt{1+\delta_k}}{\epsilon+\sigma_{1}^{2}} + 2 \right). \]
    In particular, when $x$ is $k$-sparse and $\eta=0$, then the sequence $\{x^{p}\}$ converges to $x$.
\end{theorem} \label{thm1}
\emph{Proof.}
Let $\hat{w}$ be a $k$-sparse binary vector such that $S \subseteq \operatorname{supp}(\hat{w})$, which implies $x_S = x_S \otimes \hat{w}$.
From the structure of NTOT, ${\cal Z}_k^{\#}(u^p) = u^p\otimes \overline{w}^p $ where $\overline{w}^p$ is the optimal solution to the problem (P$_1$).
Note that $y= Ax + \eta = Ax_S + \eta'$ where $\eta' = Ax_{\overline{S}} + \eta$.
By Lemma \ref{lem2}, we immediately have
\begin{align}
    \norm*{u^p \otimes \overline{w}^p - x_S}_2
    & \le \sqrt{\frac{1+\delta_k}{1-\delta_{2k}}} \norm*{(u^p-x_S) \otimes \hat{w}}_2 + \frac{2}{\sqrt{1-\delta_{2 k}}} \| \eta' \|_2. \label{eq1}
\end{align}
By the definition of $ u^p$ in NTOT, we see that
\begin{align}
u^{p}-x_{S} &=x^{p}-x_{S}+\lambda\left(A^{T} A+\epsilon I\right)^{-1} A^{T}\left(y-A x^{p}\right) \nonumber \\
&=\left(I-\lambda\left(A^{T} A+\epsilon I\right)^{-1} A^{T} A\right)\left(x^{p}-x_{S}\right)+\lambda\left(A^{T} A+\epsilon I\right)^{-1} A^{T} \eta' . \label{eq2}
\end{align}
By the singular value decomposition of $A$, for any vector $u \in \mathbb{R}^n$, it is very easy to verify that
\begin{equation} \label{tail}
	\left\|\left(A^{T} A+\epsilon I\right)^{-1} A^{T} u\right\|_{2} \leq \frac{\sigma_{1}}{\epsilon+\sigma_{1}^{2}}\|u\|_{2} .
\end{equation}
From the choices of $(\epsilon, \lambda)$, we see that $\lambda \le \epsilon+\sigma_m^2$ and $ \epsilon > \sigma_1^2$.
Thus it follows from \eqref{eq2} and \eqref{tail} that
\begin{align}
    \norm*{(u^p - x_S) \otimes \hat{w}}_2
    = & \norm*{(u^p - x_S)_{\operatorname{supp}(\hat{w})}}_2 \nonumber \\
    \le & \norm*{ \left[ (I-\lambda (A^TA + \epsilon I)^{-1}A^TA) (x^p - x_S) \right]_{\operatorname{supp} (\hat{w})} }_2 \nonumber \\
    & + \left\| \left(\lambda\left(A^{T} A+\epsilon I\right)^{-1} A^{T} \eta' \right)_{\operatorname{supp}(\hat{w})}\right\|_{2} \nonumber \\
    \le & \left( \delta_{2k} + \sigma_1^2-\frac{\lambda \sigma_1^2}{\epsilon +\sigma_1^2} \right) \norm*{x_S - x^p}_2
    + \frac{\lambda \sigma_{1}}{\epsilon+\sigma_{1}^{2}} \left\|\eta'\right\|_{2} ,\label{eq3}
\end{align}
where the first term of the right-hand side follows from Lemma \ref{lem1} with the fact $|\operatorname{supp}(\hat{w}) \cup \operatorname{supp}(x^p-x_S)| \le |\operatorname{supp}(\hat{w}) \cup S \cup S^p| \le 2k$ since $S \subseteq \operatorname{supp}(\hat{w})$.
Combining  \eqref{eq1} and \eqref{eq3} leads to
\begin{equation} \label{TTTT}
	\left\|x^{p+1}-x_S\right\|_{2}=\norm*{u^p \otimes \overline{w}^p - x_S}_2 \leq \rho \left\|x^{p}-x_S\right\|_{2} + \tau \left\|\eta'\right\|_{2} ,
\end{equation}
where
\[ \rho = \sqrt{\frac{1+\delta_{k}}{1-\delta_{2k}}} (\delta_{2k} + \sigma_1^2-\frac{\lambda \sigma_1^2}{\epsilon + \sigma_1^2}) , ~~ \tau = \frac{1}{\sqrt{1-\delta_{2 k}}}\left(\frac{\lambda \sigma_{1} \sqrt{1+\delta_{k}}}{\epsilon+\sigma_{1}^{2}}+2\right) . \]
From (\ref{TTTT}), to guarantee the recovery of $x_S$ by the NTOT, it is sufficient to ensure that $ \rho<1,$ which
 is equivalent to
\begin{equation*}
    \lambda > \epsilon + \sigma_1^2 + \left( \delta_{2k} - \sqrt{\frac{1-\delta_{2k}}{1+\delta_{k}}} \right) \frac{\epsilon + \sigma_1^2}{\sigma_1^2}.
\end{equation*}
This is guaranteed under the choice of $\lambda$ given in (\ref{LLLL}). The remaining proof is to show that the range in (\ref{LLLL}) exists.
In fact, if the following two conditions are satisfied, the existence of such a range is guaranteed:
\begin{equation} \label{eq4}
    \delta_{2k} - \sqrt{\frac{1-\delta_{2k}}{1+\delta_{k}}} < 0
\end{equation}
and
\begin{equation} \label{eq5}
     \epsilon + \sigma_1^2 + \left( \delta_{2k} - \sqrt{\frac{1-\delta_{2k}}{1+\delta_k}} \right) \frac{\epsilon + \sigma_1^2}{\sigma_1^2} < \epsilon + \sigma_m^2.
\end{equation}
By noting that $\delta_k \le \delta_{2k}$, it is straightforward to verify that the inequality \eqref{eq4} is guaranteed under the condition $\delta_{2k} < 0.5349$.
The inequality \eqref{eq5} can be written as
\[ \epsilon > \left( \frac{\sigma_1^2 - \sigma_m^2}{\sqrt{\frac{1-\delta_{2k}}{1+\delta_k}} - \delta_{2k}} - 1 \right) \sigma_1^2 , \]
which is also guaranteed under the choice of $ \epsilon$ given in \eqref{LLL}.
Thus the desired result follows.
In particular, if $\eta=0$ and $x$ is $k$-sparse, the relation \eqref{thm1result} is reduced to
\[ \left\|x^{p+1}-x\right\|_{2} \leq \rho\left\|x^{p}-x\right\|_{2} \leq \rho^p \left\|x^0-x\right\|_{2}, \]
which implies that $\{x^{p}\}$ converges to $x$ as $p \to \infty$.
\hfill $\Box$

\subsection{Analysis of NTROT in noisy scenarios}

Still we denote by $S={\cal L}_k(x)$, the index set of the largest $k$ magnitudes of $x.$
The measurements are given as $ y= Ax+ \eta$, where $ \eta \in \mathbb{R}^m $  is a noise vector.
We first recall some useful technical results which have been shown in \cite{zhao2020optimal} and \cite{zhao2020analysis}. Lemma \ref{lem3} below is a property of the hard thresholding operator $ {\cal H}_k,$ whereas the second one is property of the solution of the optimization problem (P$_2$) in NTROT and (P$_4$) in NTROTP.
\begin{lemma} \emph{\cite{zhao2020optimal}}  \label{lem3}
	Let $z \in \mathbb{R}^{n}$ be a given vector  and $v \in \mathbb{R}^{n}$ be a $k$-sparse vector with $\Phi = \operatorname{supp}(v)$.
	Denote $\Omega = {\cal L}_{k} (z)$.
	Then one has
	\[ \norm*{v-{\cal H}_{k} (z)}_{2} \le \norm*{(v-z)_{\Omega \cup \Phi}}_{2} + \norm*{(v-z)_{\Omega \backslash \Phi}}_{2}. \]
\end{lemma}

\begin{lemma} \emph{\cite{zhao2020analysis}} \label{decompose}
	Let $\Lambda \subseteq \{ 1,\dots,n \}$ be any given index set, and let $w \in \mathbb{R}^n$ be any given vector satisfying $\mathbf{e}^{T} w=k$ and $0 \leq w \leq \mathbf{e}$.
	Decompose the vector $ w_{\Lambda}$ as
	\[ w_{\Lambda} = w_{\Lambda_1} + \dots + w_{\Lambda_{{q}-1}} + w_{\Lambda_{{q}}}, \]
	where $q$ is a nonnegative integer number such that $|\Lambda|=({q}-1) k+\alpha$ where $0 \leq \alpha <k$, $w_{\Lambda_1}$ is the first $k$ largest magnitudes in $\{w_i : i \in \Lambda \}$, $w_{\Lambda_2}$ is the second $k$ largest magnitudes in $\{w_i : i \in \Lambda \}$, and so on.
	Then one has
	\[ \left\|w_{\Lambda_{1}}\right\|_{\infty}+\cdots+\left\|w_{\Lambda_{q-1}}\right\|_{\infty}+\left\|w_{\Lambda_{q}}\right\|_{\infty} \le 2 . \]
\end{lemma}

The next result is actually implied from the proof of Theorem 4.8 in \cite{zhao2020optimal}.
Item (i) in the lemma below is immediately obtained by combining two inequalities in the proof of Theorem 4.8 in \cite{zhao2020optimal}.
So we only outline a simple proof of the Item (ii) for this lemma.

\begin{lemma} \label{lem5}
Let $y=A x+\eta$ be the measurements of $x$ and $\hat{w} \in \{0,1\}^n$ be a $k$-sparse binary vector such that $S = {\cal L}_k (x) \subseteq \operatorname{supp}(\hat{w})$.
Let $S^{p+1} = \operatorname{supp}(x^{p+1})$, $u^p$ and $w^p$ be defined in NTROT.
One has
\begin{equation*}
\begin{aligned}
& \begin{aligned}
	\text{\emph{(i)}} ~ \left\|\left(x_S-u^{p} \otimes w^{p}\right)_{S \cup S^{p+1}}\right\|_{2}
	\le & \sqrt{\frac{1+\delta_{k}}{1-\delta_{2 k}}}\left\|\left(x_S-u^{p}\right) \otimes \hat{w}\right\|_{2}
	+\frac{2}{\sqrt{1-\delta_{2 k}}}\left\|\eta'\right\|_{2} \\
	& + \frac{1}{\sqrt{1-\delta_{2 k}}} \left\|A\left[\left(x_S-u^{p} \otimes w^{p}\right)_{\overline{S \cup S^{p+1}}}\right]\right\|_{2} ,
\end{aligned} \\
& \begin{aligned}
 \text{\emph{(ii)}} ~\left\|A\left[\left(x_S-u^{p} \otimes w^{p}\right)_{\overline{S \cup S^{p+1}}}\right]\right\|_{2} \le 2 \sqrt{1+\delta_k} \| {\cal H}_k (u^p-x_S) \|_2.
\end{aligned}
\end{aligned}
\end{equation*}
\end{lemma}
\emph{Proof.}
By setting $\Lambda := \overline{S \cup S^{p+1}}$ and $w := w^p$ in Lemma \ref{decompose}.
Decompose the vector $(w^p)_{\overline{S \cup S^{p+1}}}$ into
\[ (w^p)_{\overline{S \cup S^{p+1}}} = (w^p)_{\Lambda_1} + \dots +(w^p)_{\Lambda_{q-1}} + (w^p)_{\Lambda_{q}} \]
in the way described in Lemma \ref{decompose}.
Since $(x_S)_{\overline{S\cup S^{p+1}}} = 0$, we have
\begin{align*}
	\left(x_{S}-u^{p} \otimes w^{p}\right)_{\overline{S \cup S^{p+1}}}
	& = \left( (x_{S}-u^{p}) \otimes w^{p}\right)_{\overline{S \cup S^{p+1}}} \\
	& = \left( (x_{S}-u^{p}) \otimes w^{p}\right)_{\Lambda_1} + \dots + \left( (x_{S}-u^{p}) \otimes w^{p}\right)_{\Lambda_{q}} \\
	& = v^{(1)}+v^{(2)}+\cdots+v^{(q)},
\end{align*}
where $v^{(i)} = \left( (x_{S}-u^{p}) \otimes w^{p}\right)_{\Lambda_i}$, $i = 1, \dots, q$.
Thus
\[ \left\|A\left[\left(x_{S}-u^{p} \otimes w^{p}\right)_{\overline{S \cup S^{p+1}}}\right]\right\|_{2} \le \sum_{i=1}^{q}\left\|A v^{(i)}\right\|_{2} \le \sqrt{1+\delta_{k}} \sum_{i=1}^{q}\left\|v^{(i)}\right\|_{2}, \]
where the last inequality follows from the definition of $\delta_k$ and the fact $|\operatorname{supp}(v^{(i)})| \le k$.
We also have that
\begin{align*}
	\sum_{i=1}^{q} \left\|v^{(i)}\right\|_{2}
	= & \sum_{i=1}^{q} \left\|\left[\left(u^{p}-x_S\right) \otimes w^{p}\right]_{\Lambda_{i}}\right\|_{2} \\
	\le & \sum_{i=1}^{q} \left\|(w^p)_{\Lambda_i}\right\|_{\infty} \norm*{(u^p-x_S)_{\Lambda_i}}_2 \\
	\le & 2 \left\|\mathcal{H}_{k}\left(u^{p}-x_{S}\right)\right\|_{2},
\end{align*}
where the last inequality follows from the fact $\norm*{(u^p-x_S)_{\Lambda_i}}_2 \le \left\|\mathcal{H}_{k} \left(u^{p}-x_{S}\right)\right\|_{2}$ and Lemma \ref{decompose} which claims that $\sum_{i=1}^{q} \left\|(w^p)_{\Lambda_i}\right\|_{\infty} < 2$.
Combining the above two inequalities yields
\[ \left\|A\left[\left(x_{S}-u^{p} \otimes w^{p}\right)_{\overline{S \cup S^{p+1}}}\right]\right\|_{2} \le 2 \sqrt{1+\delta_{k}}\left\|\mathcal{H}_{k}\left(u^{p}-x_{S}\right)\right\|_{2} , \]
which is exactly the relation given in Item (ii) of the Lemma.
\hfill $\Box$

The main result in this section is stated as follows.
\begin{theorem} \label{thm2}
    Let $y=Ax+\eta$ be the measurements of $x \in \mathbb{R}^n$ with measurement error $ \eta.$
    Let $S={\cal L}_k(x)$ and let $\sigma_1$ and $\sigma_m$ denote, respectively, the largest and smallest singular values of the matrix $A \in \mathbb{R}^{m \times n}$.
    Suppose that the restricted isometry constant of $A$ satisfies that
    \[ \delta_{3k} < 0.2119 . \]
    Let $\epsilon$ be a given parameter satisfying
    \begin{equation} \label{thm2-1}
    	\epsilon > \max \left\{ \sigma_1^2 , \left(\frac{\sigma_{1}^{2}-\sigma_{m}^{2}}{\frac{1}{3 \sqrt{\frac{1+\delta_{3 k}}{1-\delta_{3 k}}}+1}-\delta_{3 k}}-1\right) \sigma_{1}^{2} \right\} .
    \end{equation}
    If the stepsize $\lambda$ in NTROT satisfies
    \begin{equation} \label{thm2-2}
    	\epsilon+\sigma_{1}^{2}+\left(\delta_{3 k}-\frac{1}{3 \sqrt{\frac{1+\delta_{3 k}}{1-\delta_{3 k}}+1}}\right) \frac{\epsilon+\sigma_{1}^{2}}{\sigma_{1}^{2}} ~ < ~ \lambda ~ \le ~ \epsilon+\sigma_{m}^{2},
    \end{equation}
    then the sequence $\{x^p\} $ generated by the NTROT satisfies that
    \begin{equation} \label{eq16}
    \norm*{x^{p+1} - x_{S}}_{2} \le \rho \left\|x^{p}-x_{S}\right\|_{2}+\tau\left\|A x_{\overline{S}}+\eta \right\|_{2},
    \end{equation}
    where
    \begin{equation} \label{thm2-3}
	\rho = \sqrt{\frac{1+\delta_{k}}{1-\delta_{2 k}}}\left(\delta_{2 k}+2 \delta_{3 k}+3 \sigma_{1}^{2}-3 \frac{\lambda \sigma_{1}^{2}}{\epsilon+\sigma_{1}^{2}}\right)+\delta_{3 k}+\sigma_{1}^{2}-\frac{\lambda \sigma_{1}^{2}}{\epsilon+\sigma_{1}^{2}}
	\end{equation}
    and
    \begin{equation} \label{thm2-4}
	\tau = \frac{1}{\sqrt{1-\delta_{2 k}}}\left(\frac{3 \lambda \sigma_{1} \sqrt{1+\delta_{k}}}{\epsilon+\sigma_{1}^{2}}+2\right)+\frac{\lambda \sigma_{1}}{\epsilon+\sigma_{1}^{2}}.
    \end{equation}
    In particular, when $x$ is $k$-sparse and $\eta=0$, then the sequence $\{x^{p}\}$ converges to $x$.
\end{theorem}
\emph{Proof.}
Let $ S, \sigma_1 , \sigma_m $ be defined as in the theorem. Note that $y: = Ax+\eta = Ax_{S} + \eta'$ where $\eta' = Ax_{\overline{S}} + \eta. $
Denote by $S^{p+1} = \operatorname{supp}(x^{p+1})$.
Applying Lemma \ref{lem3}, we immediately have
\begin{align}
    \norm*{x_{S}-x^{p+1}}_2 & =\norm*{x_{S}- {\cal H}_k (u^p \otimes w^p) }_2 \nonumber \\
    & \le \norm{(u^p \otimes w^p - x_{S})_{S^{p+1} \cup S}}_2 + \norm*{(u^p \otimes w^p - x_{S})_{S^{p+1} \backslash S}}_2 . \label{eq6}
\end{align}
In what follows, we bound each of the terms on the right-hand side of the above inequality.
By the definition of $ u^p$ in NTROT, we have
\begin{equation*}
	u^p-x_{S} = \left(I-\lambda\left(A^{T} A+\epsilon I\right)^{-1} A^{T} A\right)\left(x^{p}-x_S\right) + \lambda (A^{T}A + \epsilon I)^{-1} A^{T} \eta'.
\end{equation*}
Noting that $(x_{S})_{{S^{p+1} \backslash S}} = (x_{S} \otimes w^{p})_{{S^{p+1} \backslash S}} = 0$, we have
\begin{align}
    & \norm*{(u^p \otimes w^p - x_{S})_{S^{p+1} \backslash S}}_2 \nonumber \\
    & = \norm*{((u^p-x_{S}) \otimes w^p)_{S^{p+1} \backslash S}}_2 \nonumber \\
    & \le \norm*{(u^p-x_{S})_{S^{p+1} \backslash S}}_2    ~~~~~~~~~~ (\textrm{since} ~ 0 \leq w^p \leq \mathbf{e})  \nonumber \\
    & = \left\| \left[ \left(I-\lambda\left(A^{T} A+\epsilon I\right)^{-1} A^{T} A\right)\left(x^{p}-x_S\right) + \lambda (A^{T}A + \epsilon I)^{-1} A^{T} \eta' \right]_{S^{p+1} \backslash S} \right\|_{2} \nonumber \\
    & \le \left\| \left[ \left(I-\lambda\left(A^{T} A+\epsilon I\right)^{-1} A^{T} A\right)\left(x^{p}-x_S\right) \right]_{S^{p+1} \backslash S} \right\|_{2} \nonumber \\
    & \quad + \lambda \left\| \left[ (A^{T}A + \epsilon I)^{-1} A^{T} \eta' \right]_{S^{p+1} \backslash S} \right\|_{2} \nonumber \\
    & \le \left( \delta_{3k} + \sigma_{1}^{2} - \frac{\lambda \sigma_1^2}{\epsilon+\sigma_1^2} \right) \norm*{x^{p} - x_{S} }_{2}+ \frac{\lambda\sigma_1}{\epsilon + \sigma_1^2} \norm*{\eta'}_{2}, \label{eq7}
\end{align}
where the last inequality follows from Lemma \ref{lem1} (with the fact $ |\operatorname{supp}(x^p-x_S) \cup (S^{p+1}\backslash S)| \leq 3k$) and \eqref{tail}.
We now provide an upper bound for the first term of the right-hand side of \eqref{eq6}.
Let $\hat{w}$ be a $k$-sparse binary vector satisfying $S \subseteq \operatorname{supp}(\hat{w})$.
By Lemma \ref{lem5}, we have
\begin{align}
\left\|\left(x_{S}-u^{p} \otimes w^{p}\right)_{S \cup S^{p+1}}\right\|_{2}
	\le & \sqrt{\frac{1+\delta_{k}}{1-\delta_{2 k}}}\left\|\left(x_{S}-u^{p}\right) \otimes \hat{w}\right\|_{2}
	+\frac{2}{\sqrt{1-\delta_{2 k}}}\left\|\eta'\right\|_{2} \nonumber \\
	& + \frac{1}{\sqrt{1-\delta_{2 k}}} \left\|A\left[\left(x_{S}-u^{p} \otimes w^{p}\right)_{\overline{S \cup S^{p+1}}}\right]\right\|_{2} \label{eq13}
\end{align}
and
\begin{equation} \label{eq14}
	\left\|A\left[\left(x_{S}-u^{p} \otimes w^{p}\right)_{\overline{S \cup S^{p+1}}}\right]\right\|_{2} \le 2 \sqrt{1+\delta_k} \| {\cal H}_k (u^p-x_S) \|_2 .
\end{equation}
Applying Lemma \ref{lem1} (with $\epsilon > \sigma_1^2$, $\lambda \le \epsilon+\sigma_m^2$ and $|\operatorname{supp}(x^p - x_S) \cup \operatorname{supp}(\hat{w})| \le 2k$) and \eqref{tail}, we have
\begin{align}
    & \norm{(x_S - u^p) \otimes \hat{w}}_2 \nonumber \\
    & \le \norm*{ \left[ (I-\lambda (A^TA + \epsilon I)^{-1}A^TA) (x^p - x_S) \right] \otimes \hat{w} }_2 + \norm*{(\lambda (A^TA + \epsilon I)^{-1} A^T \eta') \otimes \hat{w}}_2 \nonumber \\
    & \le \left( \delta_{2k} + \sigma_1^2-\frac{\lambda \sigma_1^2}{\epsilon +\sigma_1^2} \right) \norm*{x_S - x^p}_2 + \frac{\lambda \sigma_1}{\epsilon+\sigma_1^2} \norm*{\eta'}_2. \label{eq11}
\end{align}
Denote by $\Phi = {\cal L}_k (x_{S} - u^{p})$.
As $|\operatorname{supp}(x^p-x_S) \cup \Phi | \le 3k$, by a proof similar to \eqref{eq11}, we also have
\begin{align}
    \norm*{{\cal H}_k (x_{S} - u^{p})}_2
    = & \norm*{(x_{S} - u^{p})_{\Phi}}_2 \nonumber \\
    \leq & \left(\delta_{3 k}+\sigma_{1}^{2}-\frac{\lambda \sigma_{1}^{2}}{\epsilon+\sigma_{1}^{2}}\right) \left\|x^{p}-x_{S}\right\|_{2}
    + \frac{\lambda \sigma_{1}}{\epsilon+\sigma_{1}^{2}} \norm{\eta'}_2 .\label{eq15}
\end{align}
Combining \eqref{eq13}-\eqref{eq15}, we have
\begin{equation} \label{a1}
	\left\|\left(x_{S}-u^{p} \otimes w^{p}\right)_{S \cup S^{p+1}}\right\|_{2} \le \rho' \left\|x_{S}-x^{p}\right\|_{2} + \tau' \left\|\eta^{\prime}\right\|_{2}
\end{equation}
where
\[ \rho' \coloneqq \sqrt{\frac{1+\delta_{k}}{1-\delta_{2 k}}}\left(\delta_{2 k}+2 \delta_{3 k}+3 \sigma_{1}^{2}-\frac{3\lambda \sigma_{1}^{2}}{\epsilon+\sigma_{1}^{2}}\right) \]
and
\[ \tau' = \frac{1}{\sqrt{1-\delta_{2 k}}}\left(\frac{3 \lambda \sigma_{1} \sqrt{1+\delta_{k}}}{\epsilon+\sigma_{1}^{2}}+2\right) . \]
Substituting \eqref{a1} and \eqref{eq7} into \eqref{eq6} yields \eqref{eq16} with constants $\rho$ and $\tau$ given in \eqref{thm2-3} and \eqref{thm2-4}, respectively.
Due to the fact $\delta_k \le \delta_{2k} \le \delta_{3k}$, we see from \eqref{thm2-3} that
\[ \rho \le \left( 3\sqrt{\frac{1+\delta_{3k}}{1-\delta_{3k}}} + 1 \right) \left( \delta_{3 k}+\sigma_{1}^{2}-\frac{\lambda \sigma_{1}^{2}}{\epsilon+\sigma_{1}^{2}} \right). \]
Thus to ensure $\rho < 1$, it is sufficient to require that
\[ \left( 3\sqrt{\frac{1+\delta_{3k}}{1-\delta_{3k}}} + 1 \right) \left( \delta_{3 k}+\sigma_{1}^{2}-\frac{\lambda \sigma_{1}^{2}}{\epsilon+\sigma_{1}^{2}} \right) < 1 \]
which can be written as
\[ \lambda > \epsilon + \sigma_{1}^{2} + \left( \delta_{3k} - \frac{1}{3\sqrt{\frac{1+\delta_{3k}}{1-\delta_{3k}}}+1} \right) \frac{\epsilon + \sigma_{1}^{2}}{\sigma_{1}^{2}} . \]
This together with $\lambda \le \epsilon + \sigma_m^2$ implies that if the range of $\lambda$ is given as \eqref{thm2-2}, then it guarantees that $\rho<1$.
To ensure the existence of the interval in \eqref{thm2-2}, it is sufficient to choose $\epsilon$ such that
\[ \epsilon+\sigma_{1}^{2}+\left(\delta_{3 k}-\frac{1}{3 \sqrt{\frac{1+\delta_{3 k}}{1-\delta_{3 k}}}+1}\right) \frac{\epsilon+\sigma_{1}^{2}}{\sigma_{1}^{2}} < \epsilon + \sigma_m^2, \]
which is equivalent to
\begin{equation*} \label{a2}
	\delta_{3k} - \frac{1}{3\sqrt{\frac{1+\delta_{3k}}{1-\delta_{3k}}}+1} < 0, ~~ \epsilon > \left( \frac{\sigma_{1}^{2} - \sigma_{m}^{2}}{\frac{1}{3\sqrt{\frac{1+\delta_{3k}}{1-\delta_{3k}}}+1} - \delta_{3k}} - 1 \right) \sigma_{1}^{2} .
\end{equation*}
The first condition is ensured by $\delta_{3k} < 0.2119$, and the second condition is ensured by the choice of $\epsilon$ given in \eqref{thm2-1}.
The proof of the theorem is complete.
In particular, if $\eta=0$ and $x$ is $k$-sparse, the relation \eqref{eq16} is reduced to
\[ \left\|x^{p+1}-x\right\|_{2} \leq \rho \left\|x^{p}-x\right\|_{2} \leq \rho^p \left\|x^0-x\right\|_{2}, \]
which implies that $\{x^{p}\}$ converges to $x$ as $p \to \infty$.
\hfill $\Box$

\subsection{Analysis of NTROTP in noisy scenarios}

Before showing the main result, we introduce a lemma concerning a property of the pursuit step.

\begin{lemma} \emph{\cite{zhao2020optimal}} \label{lem8}
Let $y = A \hat{x} + \nu$ be the noisy measurements of the $k$-sparse signal $\hat{x} \in \mathbb{R}^n$, and let $u \in \mathbb{R}^n$ be an arbitrary $k$-sparse vector.
Then the optimal solution of the pursuit step
\[ z^{*}=\arg \min _{z}\left\{\|y-A z\|_{2}^{2}: ~ \operatorname{supp}(z) \subseteq \operatorname{supp}(u)\right\} \]
satisfies that
\[ \left\|z^{*}-\hat{x}\right\|_{2} \leq \frac{1}{\sqrt{1-\left(\delta_{2 k}\right)^{2}}}\|\hat{x}-u\|_{2}+\frac{\sqrt{1+\delta_{k}}}{1-\delta_{2 k}}\|\nu\|_{2} . \]
\end{lemma}

\begin{theorem} \label{thm3}
Let $y=Ax+\eta$ be the measurements of the signal $x \in \mathbb{R}^n$ with measurement error $ \eta.$
    Let $S={\cal L}_k(x)$ and $\sigma_1$ and $\sigma_m$ denote, respectively, the largest and smallest singular values of the matrix $A \in \mathbb{R}^{m \times n}$.
    Suppose that the restricted isometry constant of $A$ satisfies that
\[ \delta_{3k} < 0.2 , \]
and $\epsilon$ is a given parameter satisfying
\begin{equation} \label{thm3-1}
	\epsilon > \max \left\{ \sigma_1^2 , \left( \frac{\sigma_{1}^{2} - \sigma_{m}^{2}}{\frac{1}{\frac{3}{1-\delta_{3k}} + \frac{1}{\sqrt{1-\left(\delta_{3 k}\right)^{2}}}} - \delta_{3 k}} - 1 \right) \sigma_1^2 \right\} .
\end{equation}
If the stepsize $\lambda$ in NTROTP satisfies
\begin{equation} \label{thm3-2}
	\epsilon+\sigma_{1}^{2} + \left( \delta_{3 k} - \frac{1}{\frac{3}{1-\delta_{3k}} + \frac{1}{\sqrt{1-\left(\delta_{3 k}\right)^{2}}}} \right) \frac{\epsilon + \sigma_1^2}{\sigma_1^2} ~ < ~ \lambda ~ \le ~ \epsilon + \sigma_m^2 ,
\end{equation}
then the sequence $\{x^p\}$ generated by the NTROTP satisfies that
\begin{equation} \label{rhotau}
	\left\|x^{p+1}-x_{S}\right\|_{2} \leq \tilde{\rho} \left\|x^{p}-x_{S}\right\|_{2}+\tilde{\tau} \left\|A x_{\overline{S}}+\eta\right\|_{2}
\end{equation}
where
\begin{equation}
	\tilde{\rho} = \left( \frac{3}{1-\delta_{3k}} + \frac{1}{\sqrt{1-\left(\delta_{3 k}\right)^{2}}} \right) \left( \delta_{3 k}+\sigma_{1}^{2}-\frac{\lambda \sigma_{1}^{2}}{\epsilon+\sigma_{1}^{2}} \right)  \label{rho}
\end{equation}
and
\begin{equation}
	\tilde{\tau} = \frac{\sqrt{1+\delta_{k}}}{1-\delta_{2 k}} + \frac{1}{(1-\delta_{2k}) \sqrt{1+\delta_{2k}}} \left(\frac{3 \lambda \sigma_{1} \sqrt{1+\delta_{k}}}{\epsilon+\sigma_{1}^{2}}+2\right) + \frac{\lambda \sigma_{1}}{(\epsilon+\sigma_{1}^{2}) \sqrt{1-\left(\delta_{2 k}\right)^{2}}} . \label{tau}
\end{equation}
In particular, when $x$ is $k$-sparse and $\eta=0$, then the sequence $\{x^{p}\}$ converges to $x$.
\end{theorem}

\emph{Proof.}
NTROTP comprises of NTROT and a pursuit step.
From the proof of Theorem 2, we see that
\begin{equation} \label{eq19}
	\left\|x_{S}-{\cal H}_k (u^p \otimes w^p)\right\|_{2} \leq \rho\left\|x^{p}-x_{S}\right\|_{2}+\tau \left\|\eta^{\prime}\right\|_{2}
\end{equation}
where the constants $\rho$ and $\tau$ are given by \eqref{thm2-3} and \eqref{thm2-4} respectively.
From the step (P$_5$), $x^{p+1}$ is the solution to the pursuit step.
By Lemma \ref{lem8}, we have
\begin{equation} \label{eq20}
	\left\|x_{S}-x^{p+1}\right\|_{2} \le \frac{1}{\sqrt{1-\left(\delta_{2 k}\right)^{2}}} \left\|x_{S}-{\cal H}_k (u^p \otimes w^p)\right\|_{2} + \frac{\sqrt{1+\delta_{k}}}{1-\delta_{2 k}}\|\eta'\|_{2} ,
\end{equation}
where $\eta' = Ax_{\overline{S}}+\eta$.
Using $\delta_{k} \le \delta_{2k} \le \delta_{3k}$ and combining \eqref{eq19} and \eqref{eq20} leads to
\begin{equation*}
	\left\|x_{S}-x^{p+1}\right\|_{2} \leq \tilde{\rho} \left\|x^{p}-x_{S}\right\|_{2}+\tilde{\tau} \left\|\eta'\right\|_{2},
\end{equation*}
where $\tilde{\rho}$ and $\tilde{\tau}$ are defined as \eqref{rho} and \eqref{tau} respectively.
Note that $\tilde{\rho} < 1$ is equivalent to
\[ \lambda > \epsilon+\sigma_{1}^{2} + \left( \delta_{3 k} - \frac{1}{\frac{3}{1-\delta_{3k}} + \frac{1}{\sqrt{1-\left(\delta_{3 k}\right)^{2}}}} \right) \frac{\epsilon + \sigma_1^2}{\sigma_1^2}. \]
This is ensured by the choice of $\lambda$ given in \eqref{thm3-2}.
This means the choice of $\lambda$ in \eqref{thm3-2} ensures that $\tilde{\rho}<1$.
To guarantee the existence of the range \eqref{thm3-2}, it is sufficient to require that
\begin{equation} \label{eq17}
	\epsilon+\sigma_{1}^{2} + \left( \delta_{3 k} - \frac{1}{\frac{3}{1-\delta_{3k}} + \frac{1}{\sqrt{1-\left(\delta_{3 k}\right)^{2}}}} \right) \frac{\epsilon + \sigma_1^2}{\sigma_1^2} < \epsilon + \sigma_m^2 ,
\end{equation}
which is equivalent to
\begin{equation} \label{eq18}
	\delta_{3 k} - \frac{1}{\frac{3}{1-\delta_{3k}} + \frac{1}{\sqrt{1-\left(\delta_{3 k}\right)^{2}}}} < 0, ~~ \epsilon > \left( \frac{\sigma_{1}^{2} - \sigma_{m}^{2}}{\frac{1}{\frac{3}{1-\delta_{3k}} + \frac{1}{\sqrt{1-\left(\delta_{3 k}\right)^{2}}}} - \delta_{3 k}} - 1 \right) \sigma_1^2 .
\end{equation}
Note that $\frac{1}{\sqrt{1-\left(\delta_{3 k}\right)^{2}}}< \frac{1}{1-\delta_{3k}}$.
The first condition in \eqref{eq18} is satisfied when $\delta_{3k}<0.2$.
The second condition \eqref{eq18} is also satisfied provided $\epsilon$ is chosen large enough, i.e., satisfying \eqref{thm3-1}.
In particular, if $\eta=0$ and $x$ is $k$-sparse, the relation \eqref{rhotau} is reduced to
\[ \left\|x^{p+1}-x\right\|_{2} \leq \tilde{\rho}\left\|x^{p}-x\right\|_{2} \leq \tilde{\rho}^p \left\|x^0-x\right\|_{2}, \]
which implies that $\{x^{p}\}$ converges to $x$ as $p \to \infty$.
\hfill $\Box$

\section{Numerical Experiments}
Simulations were performed to test the performance of the proposed algorithms with respect to residual reduction, average number of iterations needed for convergence and success frequency for signal recovery.
The measurement matrices generated for experiments are Gaussian random matrices, whose entries are independent and identically distributed and follow the standard normal distribution ${\cal N}(0,1)$.
Nonzero entries of realized sparse signals also follow the ${\cal N}(0,1)$, and their position follows a uniform distribution.
All involved optimization problems in algorithms were solved by the CVX which is developed by Grant and Boyd \cite{cvx2017grant} with solver `Mosek'.

\subsection{Residual reduction}

\begin{figure}
\centering
\subfigure[]{
\includegraphics[width=5.5cm]{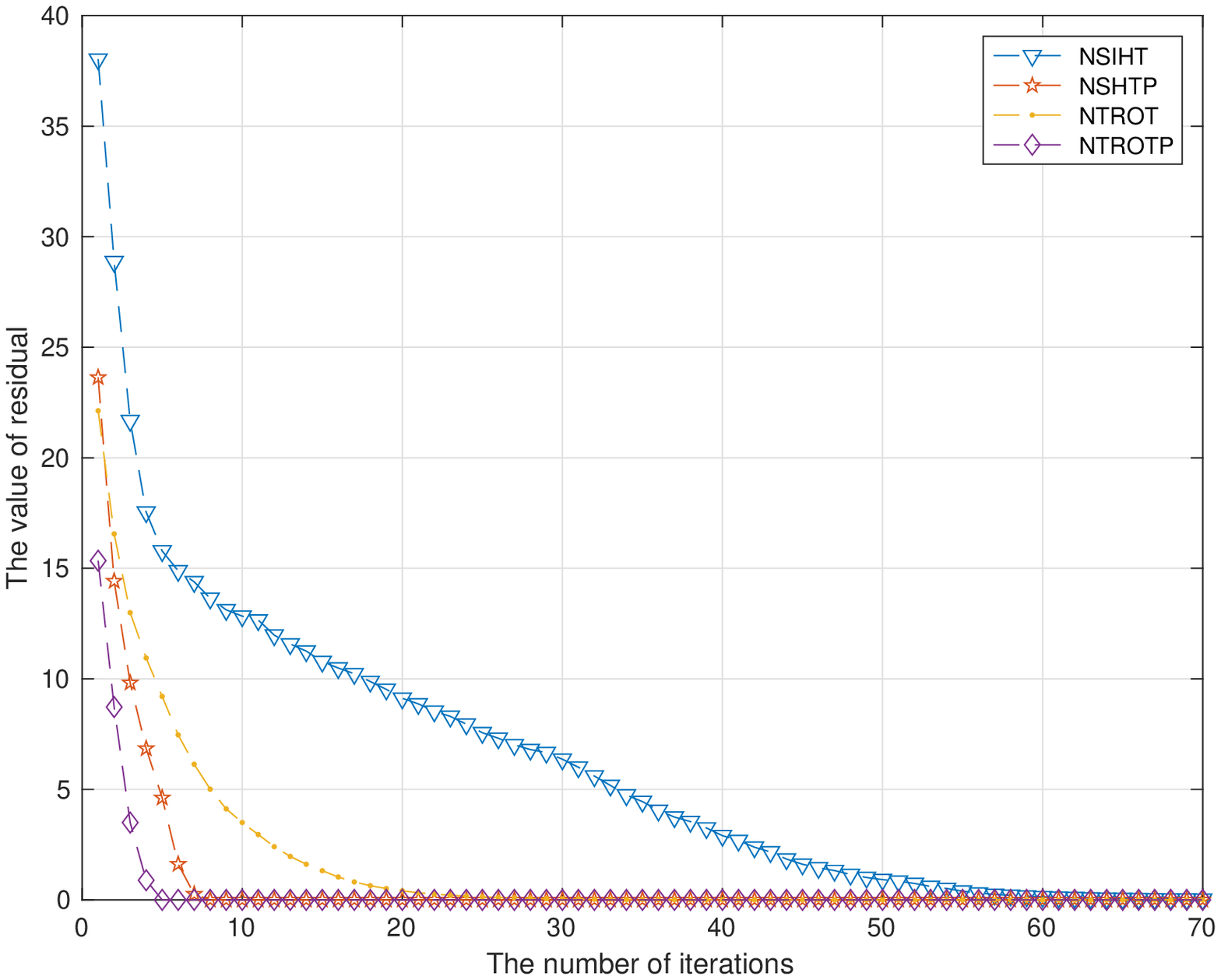}}
\subfigure[]{
\includegraphics[width=5.5cm]{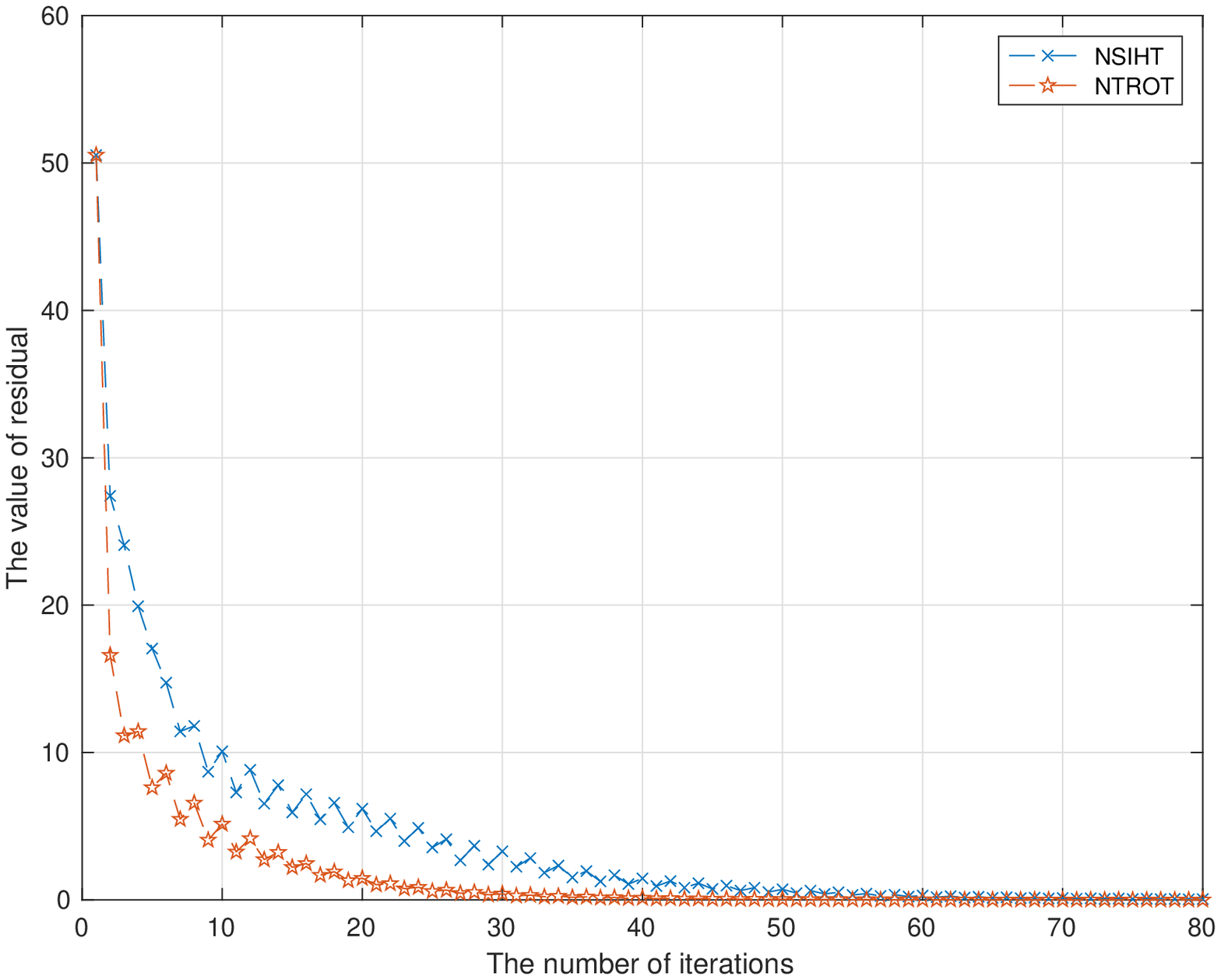}}
\caption{(a) Comparison of residual-reduction performances of several algorithms; (b) Residual change in the course of iterations using different thresholding operators}
\label{graph1}
\end{figure}

The experiment was carried out to compare the residual-reduction performance of the algorithms with given $(\epsilon, \lambda)$.
In this experiment, we set $A \in \mathbb{R}^{256 \times 512}$, $y = Ax^*$, $\|x^*\|_0 = 70$ and $x^0=0$.
The stepsize $\lambda$ and parameter $\epsilon$ are set respectively as
\begin{equation} \label{eq5-1}
	\lambda = 5, ~\epsilon = \max \{ \sigma_1^2+1, \lambda-\sigma_m^2 \},
\end{equation}
which guarantees that $\epsilon > \sigma_1^2$ and $\lambda \le \epsilon + \sigma_m^2$, where $\sigma_1$ and $\sigma_m$ denote the largest and the smallest singular value of the matrix.
Fig. \ref{graph1} demonstrates the change of the residual value, i.e., $\norm{y-Ax}_2$, in the course of iterations of the algorithms.
From Fig. \ref{graph1} (a), it can be seen that the NTROTP is more powerful than other algorithms in residual reduction.
In the same experiment environment, we also compare the residual change in the course of NSIHT and NTROT which use different thresholding operators.
Fig. \ref{graph1} (b) shows that the algorithm with optimal thresholding operator can reduce the residual more efficiently than the one with hard thresholding operator.

\begin{figure}
\centering
\subfigure[NTROT with different $\epsilon$]{
\includegraphics[width=5.5cm]{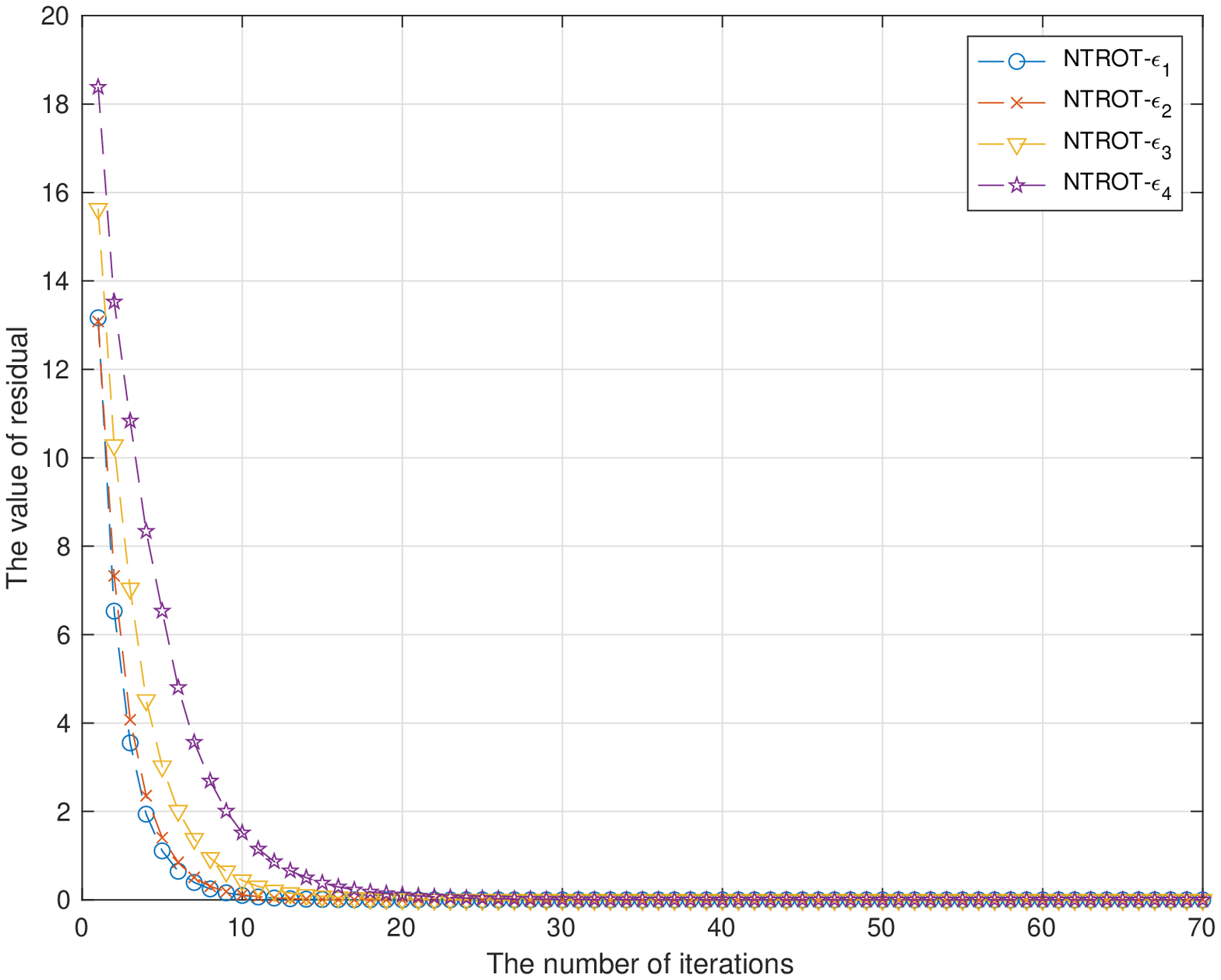}}
\subfigure[NTROTP with different $\epsilon$]{
\includegraphics[width=5.5cm]{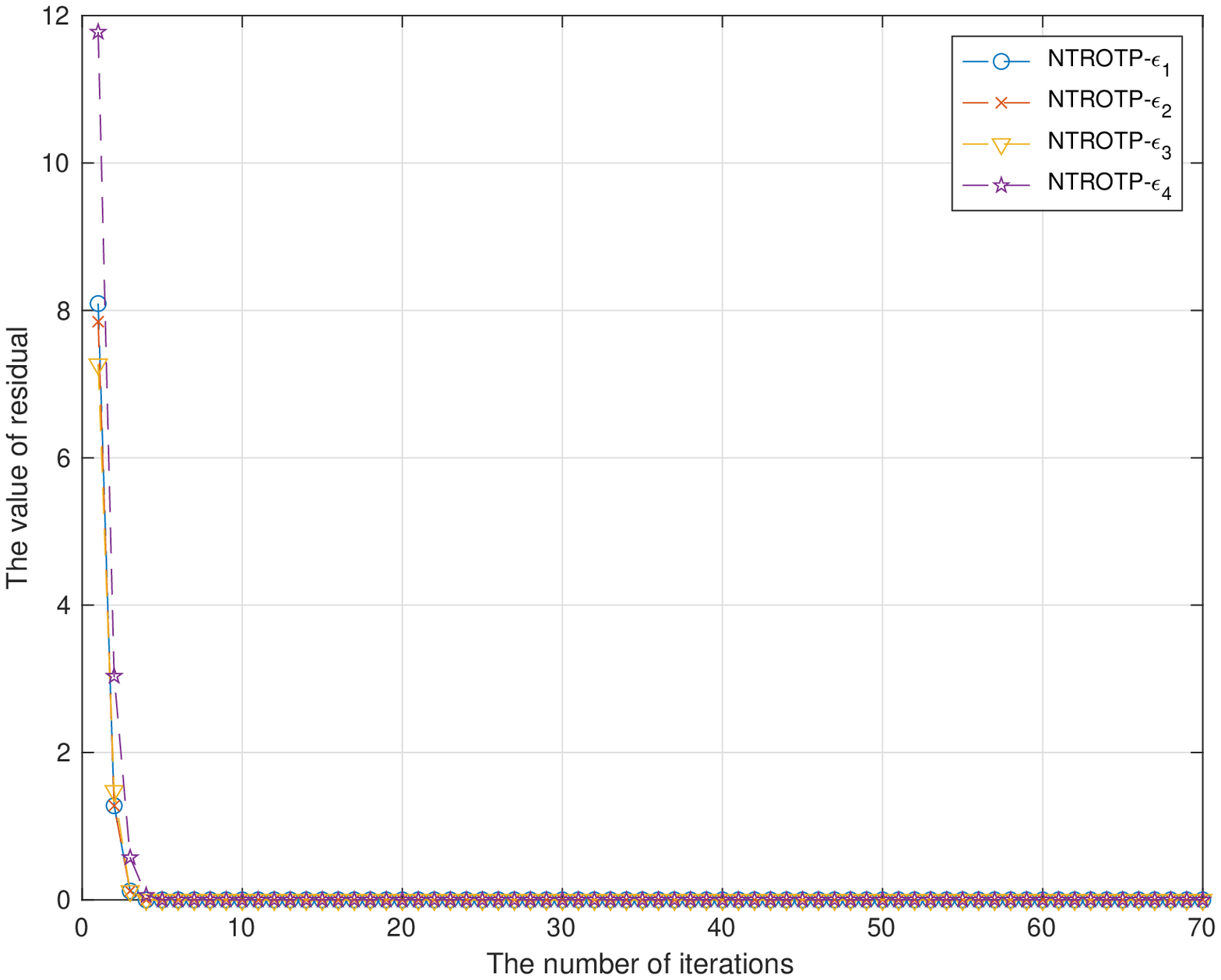}}
\caption{Residual reduction by NTROT and NTROTP with $\lambda=10$ and different parameter $\epsilon$: $\epsilon_1=\epsilon^*$, $\epsilon_2 = 1.1 \epsilon^*$, $\epsilon_3 = 1.5 \epsilon^*$, $\epsilon_4 = 2 \epsilon^*$. }
\label{graph2}
\end{figure}

\begin{figure}
\centering
\subfigure[NTROT with different $\lambda$]{
\includegraphics[width=5.5cm]{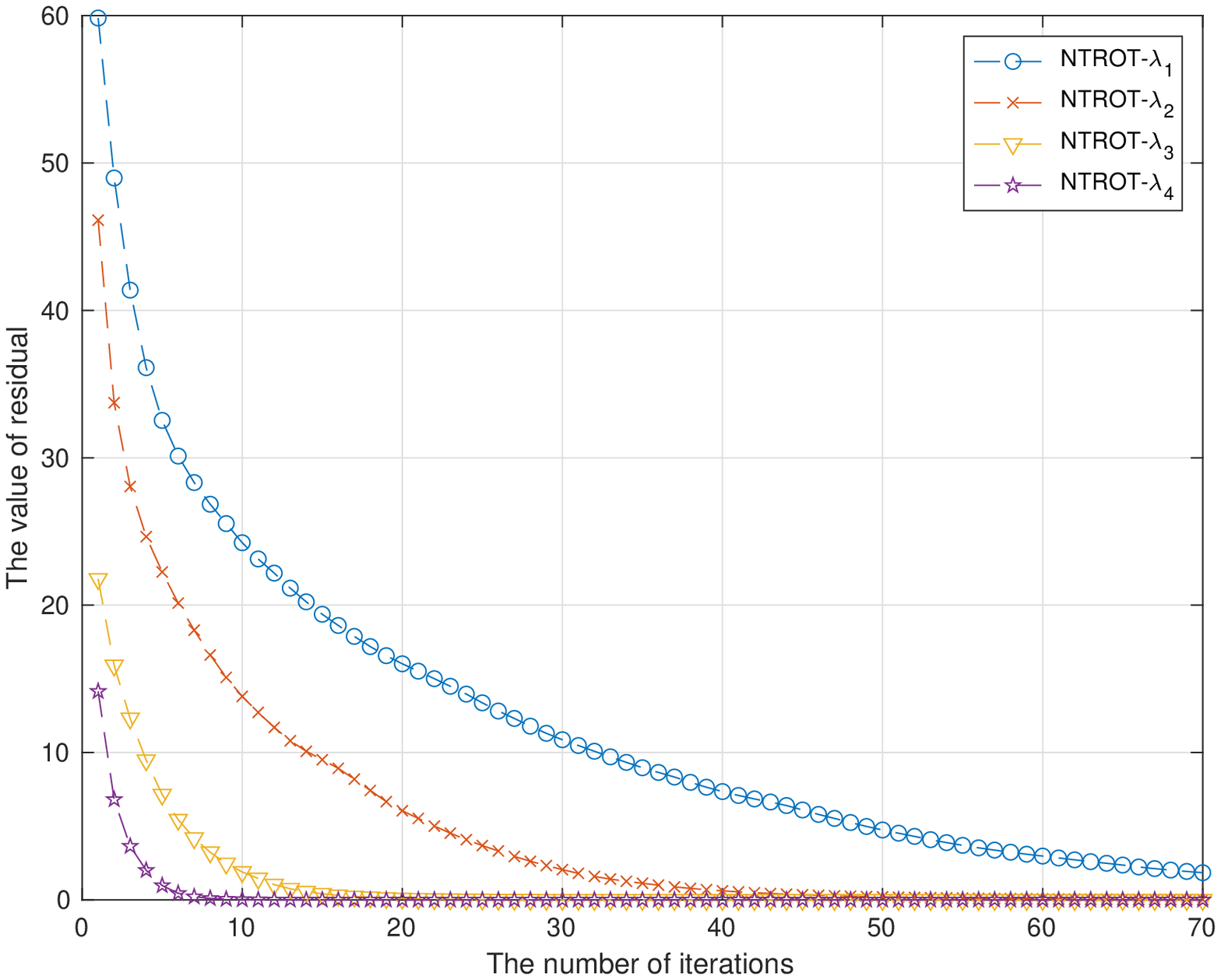}}
\subfigure[NTROTP with different $\lambda$]{
\includegraphics[width=5.5cm]{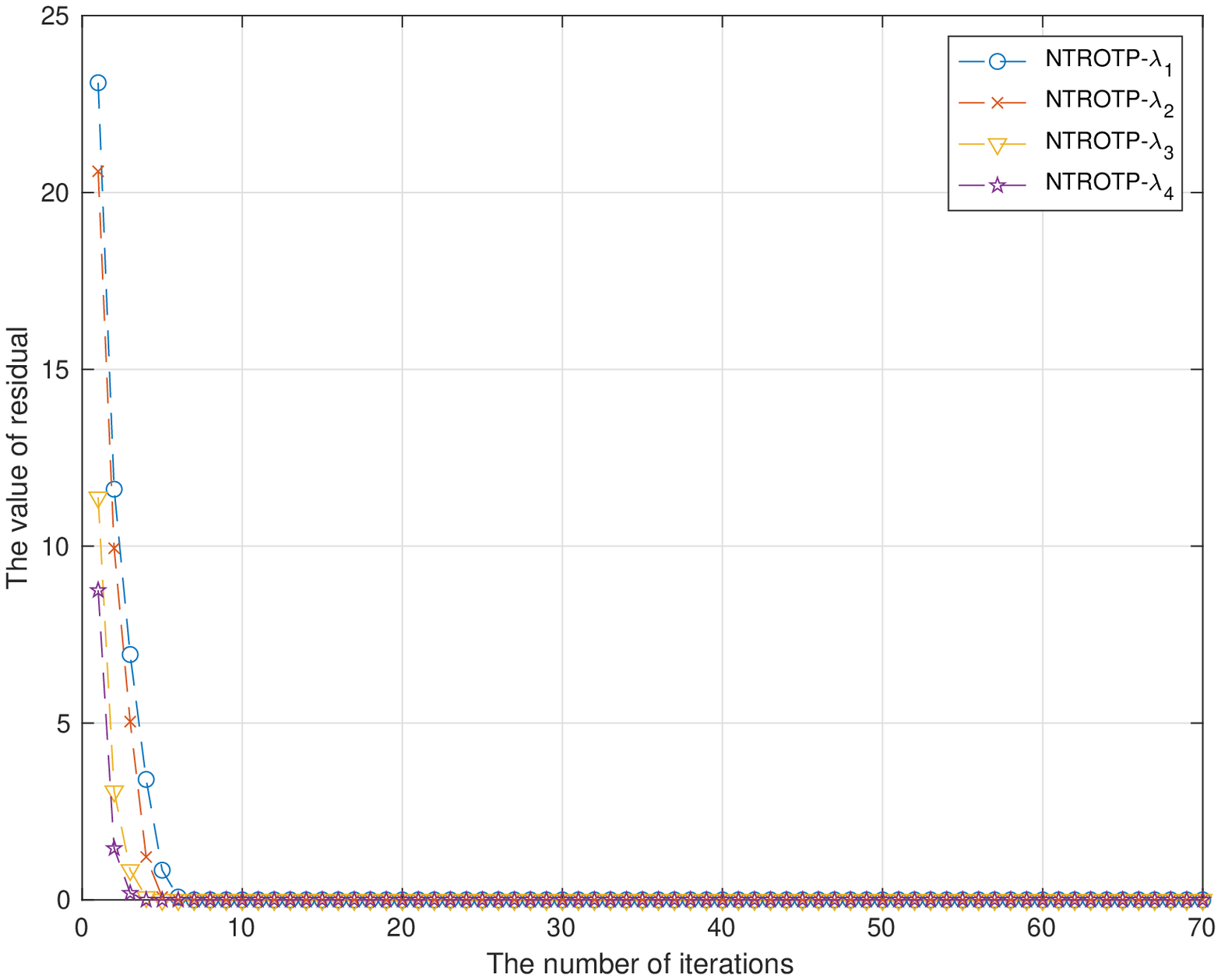}}
\caption{Residual reduction by NTROT and NTROTP with $\epsilon = \sigma_1^2+1$ and different stepsize $\lambda = 1, 2, 5, 10$.}
\label{graph3}
\end{figure}

The performance of NTROT and NTROTP are clearly related to the choice of $(\epsilon, \lambda)$.
Thus we test the residual-reduction performance of the proposed algorithms in terms of different values of parameter $\epsilon$ and stepsize $\lambda$.
The results are shown in Fig. \ref{graph2} and Fig. \ref{graph3}, respectively.
In Fig. \ref{graph2}, the stepsize $\lambda$ is fixed as $\lambda=10$, and $\epsilon = \epsilon^*, 1.1\epsilon^*, 1.5 \epsilon^*$ and $2\epsilon^*$, where $\epsilon^* = \sigma_1^2+1$.
In Fig. \ref{graph3}, the parameter $\epsilon$ is fixed as $\epsilon = \sigma_1^2+1$, and stepsize $\lambda$ is taken as $\lambda = 1,2,5,10$ respectively.
Such choices of $(\epsilon, \lambda)$ satisfy that $\epsilon>\sigma_{1}^{2}$ and $\lambda \leq \epsilon+\sigma_{m}^{2}$.
It can be seen that the NTROT is more sensitive to the change of $\epsilon$ and $\lambda$ than the NTROTP which is generally insensitive to the change of $(\epsilon, \lambda)$.
This indicates that NTROTP is a stable algorithm.

\subsection{Number of iterations}

The simulations were also performed to examine the impact of sparsity levels and measurement levels on the average number of iterations needed for signal reconstruction via Newton-type iterative algorithms.
In this experiment, all algorithms start from $x^0=0$ and terminate either when $r := \left\|x^{p}-x^{*}\right\|_{2} /\left\|x^{*}\right\|_{2} \leq 10^{-3}$ is met or when the maximum number of iterations (i.e., 50 iterations) is reached.

\begin{figure}
\centering
\subfigure[Different sparsity levels]{
\includegraphics[width=5.5cm]{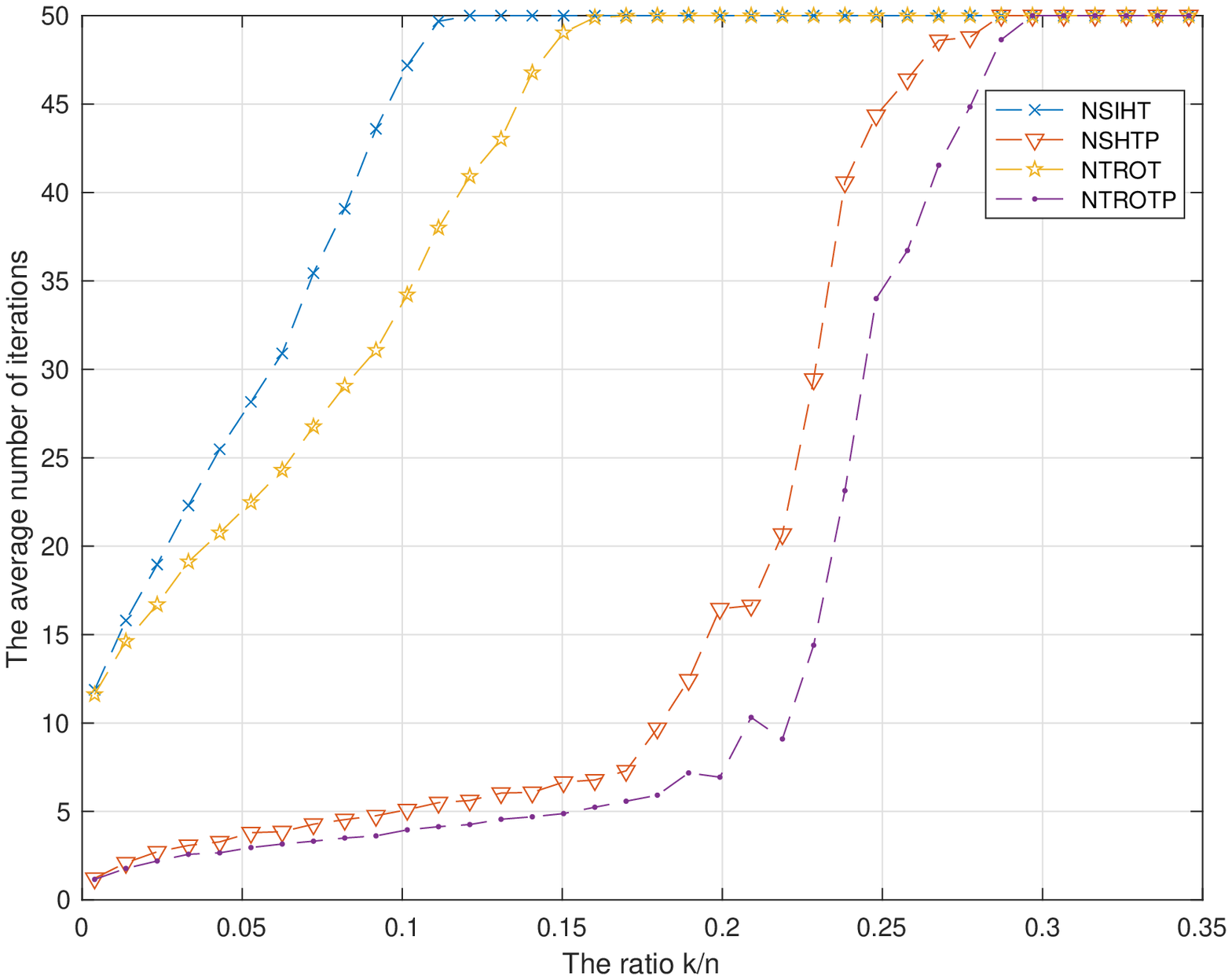}}
\subfigure[Different measurement levels]{
\includegraphics[width=5.5cm]{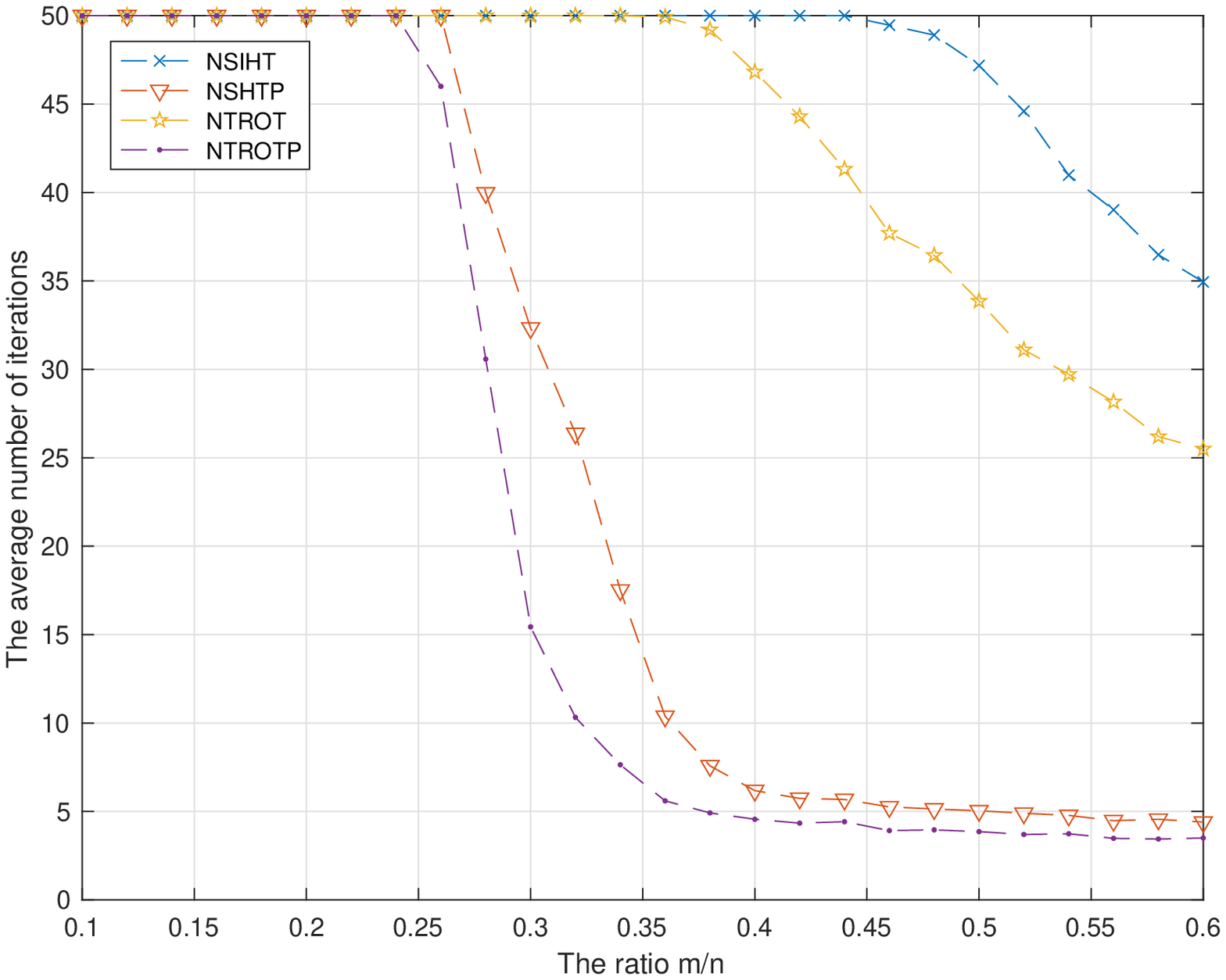}}
\caption{Comparison of the average number of iterations required by different algorithms}
\label{graph5}
\end{figure}

Fig. \ref{graph5} (a) demonstrates the influence of sparsity levels on the number of iterations needed by NSIHT, NSHTP, NTROT and NTROTP to reconstruct a signal.
In this experiment, the size of measurement matrices is still $256 \times 512$, and the ratio $k/n$ varies from 0.01 to 0.35.
The average number of iterations is calculated based on 50 random examples for each sparsity level $k/n$.
A common feature of these algorithms is that with increase of the sparsity levels, the required iterations for the algorithms to reconstruct signals also increase.
We also observe that both the optimal thresholding and pursuit step help reduce the required number of iterations of algorithms to reconstruct a signal.

Fig. \ref{graph5} (b) compares the average number of iterations required by several algorithms applying to different measurement levels.
The target signal is fixed as $x^* \in \mathbb{R}^{500}$ with $\norm{x^*}_0 = 50$, and the length of observed vector $y = Ax^*$, i.e., the number of measurements, varies from 50 to 300.
When $m/n < 0.25$, we see that no algorithm could recover the target 50-sparse signal within 50 iterations, due to the fact that the measurement levels are too low for signal reconstruction.
The more measurements obtained for the target signal $x^*$, the less number of  iterations needed for reconstruction, as shown in Fig. \ref{graph5} (b).
Both NSHTP and NTROTP could recover the signal by using relatively a small number of iterations when the ratio $m/n\ge0.35$, and the NTROTP needs less iterations than NSIHT, NSHTP and NTROT.

\subsection{Performance of signal recovery}

\begin{figure}
\centering
\subfigure[Under exact measurements]{
\includegraphics[width=5.5cm]{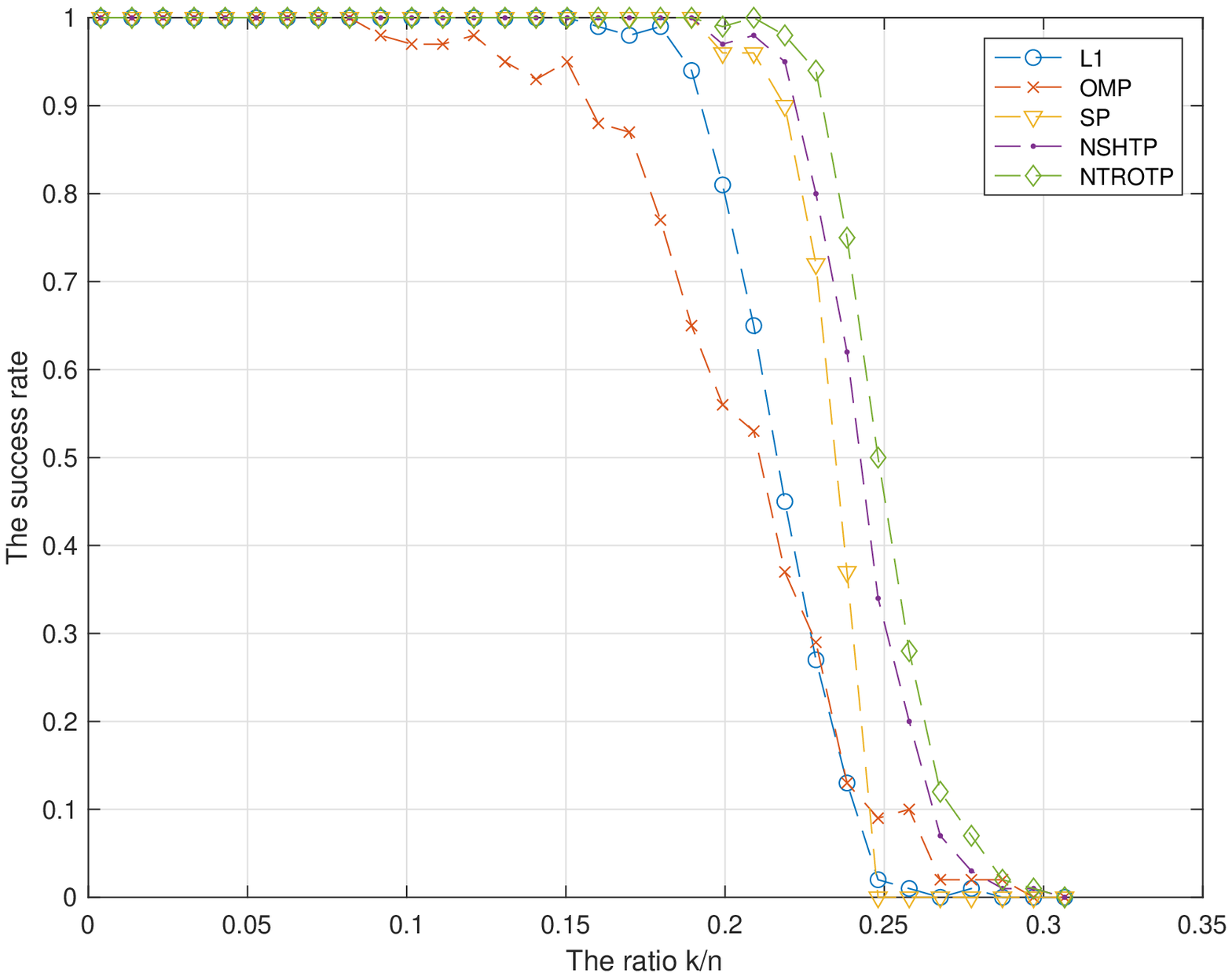}}
\subfigure[Under inexact measurements]{
\includegraphics[width=5.5cm]{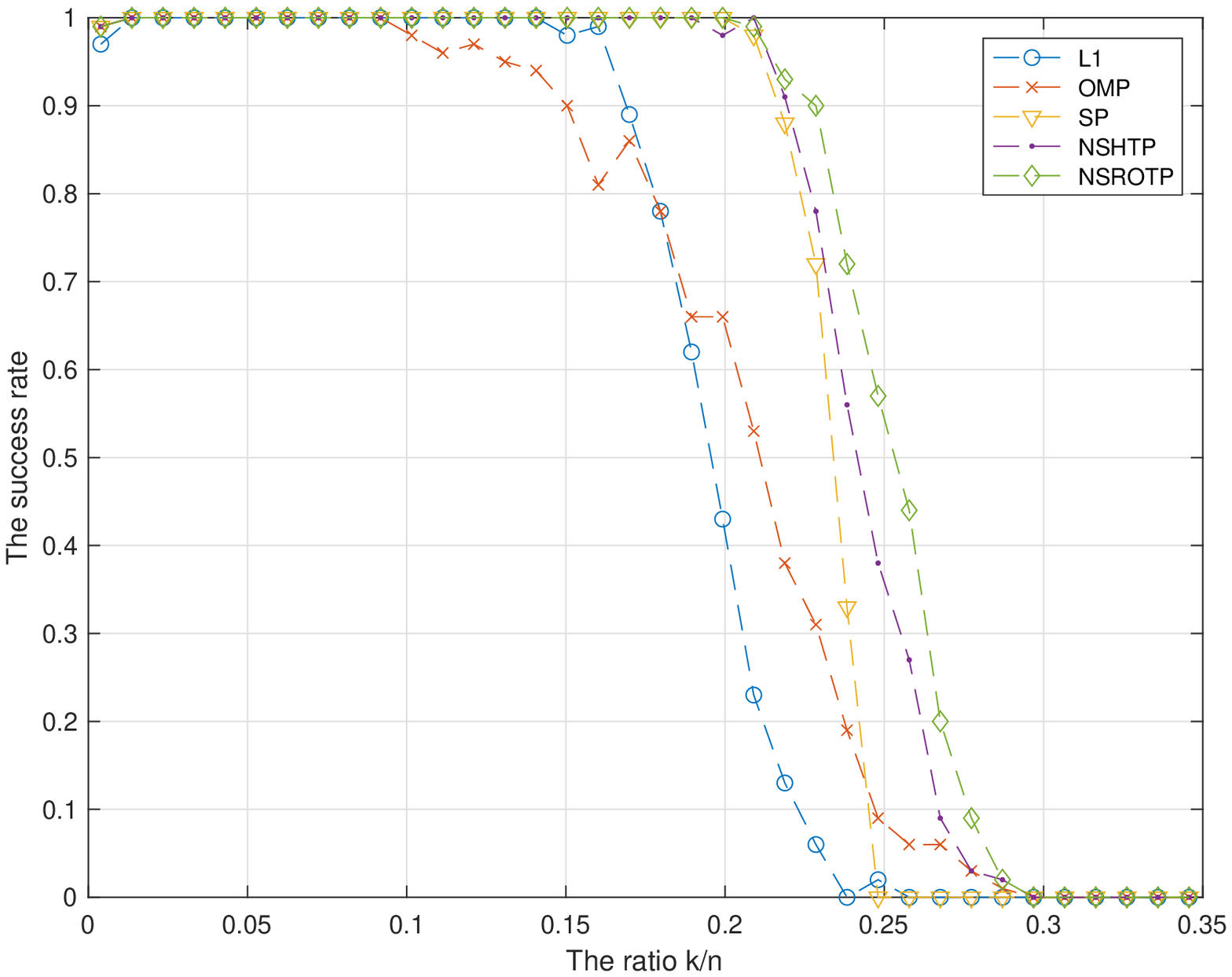}}
\caption{Comparison of success frequencies of signal recovery via different algorithms}
\label{graph6}
\end{figure}

Fig. \ref{graph6} compares the success frequencies of signal reconstruction via several algorithms with both exact and inexact measurements.
Five algorithms are taken into account in this comparison, including $\ell_1$-minimization, orthogonal matching pursuit (OMP), subspace pursuit (SP), NSHTP and NTROTP.
The size of matrices is still $256 \times 512$.
For noisy case, the measurements of $x^*$ are set as $y=Ax^*+0.001\theta$ where $\theta \in \mathbb{R}^{256}$ is a standard Gaussian random vector.
Iterative algorithms start at $x^0=0$ and terminate after 20 iterations except for the OMP which stops after $k$ iterations owing to the structure of algorithm, where $k = \norm{x^*}_0$.
The choice of $(\epsilon, \lambda)$ is the same as \eqref{eq5-1}.
The ratio $k/n$ increases from 0.01 to 0.35.
The condition $\norm{x^p-x^*}_2/\norm{x^*}_2 \le 10^{-3}$ is set as the recovery criterion.
The vertical axis in Fig. \ref{graph6} represents the success rate of reconstruction, which is calculated based on 50 random examples.
The results in Fig. \ref{graph6} indicate that the NTROTP is stable and robust for sparse signal recovery compared with other algorithms used in this experiments.

\section{Conclusion}

A class of Newton-type optimal $k$-thresholding algorithms is proposed in this paper.
Under the restricted isometry property (RIP), we have proved that the NTOT, NTROT and NTROTP algorithms are guaranteed to reconstruct sparse signals with a proper choice of the algorithmic parameter and iterative stepsize.
Simulations indicate that the proposed algorithms especially the NTROTP is a stable and robust algorithm for signal reconstruction.




\end{document}